\documentclass[draft]{agujournal2019}
\usepackage{url} 
\usepackage{lineno}
\usepackage[inline]{trackchanges} 
\usepackage{soul}

\usepackage[T1]{fontenc}
\usepackage{amsmath,amsfonts,amssymb,amsthm}
\usepackage{fix-cm}
\usepackage{microtype}
\usepackage{nicefrac}
\draftfalse

\journalname{Geophysical Research Letters}

\begin{document}

%
%


\title{Evolution of lunar wake potentials: structure, energy conversion, and their imprints on velocity distributions}

%
%




\authors{Xin An\affil{1,2}, Vassilis Angelopoulos\affil{2}, Jasper~S.~Halekas\affil{3}, Terry Z. Liu\affil{4}, Shaosui Xu\affil{5}, Andrew~R.~Poppe\affil{5}, Ferdinand Plaschke\affil{6}}


\affiliation{1}{Department of Space Science, University of Alabama in Huntsville, Huntsville, AL, 35805, USA}
\affiliation{2}{Department of Earth, Planetary, and Space Sciences, University of California, Los Angeles, CA, 90095, USA}
\affiliation{3}{Department of Physics and Astronomy, University of Iowa, Iowa City, IA, 52242, USA}
\affiliation{4}{Shandong Key Laboratory of Space Environment and Exploration Technology, Institute of Space Sciences, School of Space Science and Technology, Shandong University, Shandong, China}
\affiliation{5}{Space Sciences Laboratory, University of California, Berkeley, Berkeley, CA, 94720, USA}
\affiliation{6}{Institute of Geophysics and Extraterrestrial Physics, Technische Universit\"at Braunschweig, Braunschweig, Germany}




\correspondingauthor{Xin An}{phyax1@gmail.com}



\begin{keypoints}
\item Lunar wake electric potentials have two spatial scales: a macroscale from plasma expansion and a microscale from ion acoustic shocks.
\item Macroscale potential converts electron thermal energy to ion kinetic energy; microscale potential converts it back to thermal energy.
\item Each potential type imprints a unique signature on particle velocity distributions: ion beams and electron flat-top distributions.
\end{keypoints}

%
%

%
%


\begin{abstract}
We study the evolution of electric potentials in the lunar wake. The wake potential exhibits two distinct spatial scales. The macroscopic scale arises from solar wind expansion into the vacuum, with a potential length-scale growing with distance from the Moon; the microscopic scales arises from ion acoustic shocks near the wake center, with transition layers spanning tens of local Debye lengths. This two-scale potential mediates energy conversion between ions and electrons during wake refilling. The macroscale potential retards electrons and accelerates ions to supersonic velocities, converting electron thermal energy to ion kinetic energy. The microscale potential then decelerates ions to subsonic velocities and heats both species, converting ion kinetic energy back to thermal energy. Together, the two-scale potential imprints distinct signatures on velocity distributions, including ion beams and electron flat-top distributions, consistent with ARTEMIS observations.
\end{abstract}

\section*{Plain Language Summary}
When the Moon travels through the solar wind it blocks this flow and creates a region of near-empty space behind it, much like the wake behind a boat. This empty region is gradually refilled by the surrounding plasma, a process driven by electric fields that build up spontaneously as the faster electrons rush in ahead of the slower ions. Using computer simulations and spacecraft observations, we find that these electric fields have two distinct sizes. Large-scale electric fields, stretching over distances comparable to the Moon itself, slow down electrons and speed up ions to supersonic velocities, converting heat carried by electrons into directed motion of ions. Smaller-scale electric fields, concentrated in narrow layers near the center of the wake, then slow the ions back down and heat both ions and electrons. Together, these two types of electric fields act as an energy relay, shuffling energy between different particle populations as the wake refills. We also show that these electric fields leave recognizable fingerprints on the way electrons move, which spacecraft can detect and use to measure the electric fields indirectly, a valuable tool since direct measurement of these weak electric fields remains beyond the capability of current instruments.

%
%

%


%
%
%
%


\section{Introduction}
The Moon lacks both a global atmosphere and an intrinsic magnetic field, so solar wind particles are largely absorbed by the dayside surface, leaving a plasma void downstream \cite{holmstrom2012interaction,halekas2015moon}. This void is gradually refilled by the solar wind, forming an elongated cavity known as the lunar wake. The refilling process depends on the orientation of the interplanetary magnetic field (IMF) relative to the solar wind-void boundary normal. When the two are perpendicular, ions with large gyroradii act as a piston, driving electromagnetic plasma transport across field lines at the magnetosonic speed \cite{fatemi2013lunar,zhang2016alfven}. When the two are parallel, the faster electrons lead ions into the wake, generating an ambipolar electric field that retards electrons while accelerating ions; this electrostatic process operates at the ion acoustic speed \cite{farrell1997electrostatic,birch2002two}. Under typical IMF orientations at lunar orbit, perpendicular refilling likely dominates in the north and south sectors of the wake, while parallel refilling prevails in the dawn and dusk sectors, where most spacecraft measurements are made. Here we focus on the parallel refilling process.

The initial stage of lunar wake refilling is a classical problem of plasma expansion into vacuum, sharing physics with other contexts including space shuttle wakes \cite{tribble1988large,farrell2002similarities}, laser-solid interactions \cite{mora2003plasma}, and Mach probe wakes in laboratory plasmas \cite{hutchinson2008ion,hutchinson2008oblique}. Early theoretical work obtained a self-similar solution: an exponential plasma density decay into the vacuum and a rarefaction wave propagating back into the unperturbed plasma at the ion acoustic speed \cite<e.g.,>[]{gurevich1966self,samir1983expansion}. In this solution, the ambipolar electric field generated by charge separation accelerates ions to supersonic velocities while retarding electrons. Observational confirmation came from the Wind spacecraft, which detected counter-propagating supersonic ion beams in the central wake during two lunar flybys \cite{ogilvie1996observations}, as well as broadband electrostatic noise identified as ion acoustic waves \cite{bale1997evidence,kellogg1996observations}. These observations motivated particle-in-cell (PIC) simulations of the kinetic lunar wake \cite{farrell1998simple,birch2001detailed,birch2001particle,birch2002two}, which examined ion-ion two-stream instabilities and the resulting ion acoustic turbulence in the central wake. However, decelerating the supersonic ion beams to subsonic velocities requires ion acoustic shocks in addition to ion acoustic waves \cite{An2025plasma}; such shocks have recently been detected by ARTEMIS \cite{liu2025artemis}.

Electric potentials in the lunar wake have been inferred from electron measurements by multiple spacecraft. Using a database of more than $6000$ near-circular lunar polar orbits from the Lunar Prospector Magnetometer/Electron Reflectometer instrument, \citeA{halekas2005electrons} constructed a statistical map of electron density consistent with the self-similar expansion solution, and systematically inferred wake potentials from the shift in electron energy distributions between the solar wind and the wake. \citeA{xu2019mapping} extended this approach to ARTEMIS electron measurements, mapping wake potentials out to $4$ lunar radii downstream. These studies consistently showed a potential drop of $\sim 300$\,V in the central wake immediately downstream of the Moon, decreasing with distance as solar wind refilling progresses, reaching $\sim 50$\,V at $4$ lunar radii. However, these inferences assume that the electric potential asymmetry arises solely from the IMF orientation, and do not account for the shock region in the central wake, where electrons originating from both sides mix and interpenetrate. The microscale potential enhancements associated with ion acoustic shocks are therefore not captured by existing statistical maps, leaving their observational characterization as an open problem. This is the gap addressed by the present study.

In this study, we address the open question regarding the structure and consequences of lunar wake potentials using PIC simulations, substantiated by ARTEMIS observations. (1) We provide an evolutionary picture of lunar wake potentials that captures both the macroscale potential well associated with plasma expansion and the microscale potential enhancements associated with ion acoustic shocks in a single coherent description. (2) We verify that electrons are in force balance throughout the wake (i.e., the electric force on electrons is balanced by the electron pressure gradient), providing simulation support for the quasi-static electron response assumption underlying existing potential inference methods.

(3) We quantify how the two-scale potential mediates energy conversion between ions and electrons, exploiting the complete energy budget available in PIC simulations to trace the conversion between bulk kinetic and thermal energies at both spatial scales. (4) We characterize the imprints of the two-scale potential on particle velocity distributions, identifying ion beam and electron flat-top signatures that can be exploited to infer wake potentials from spacecraft measurements. The simulation results for the two-scale potential structure, energy conversion, and particle velocity distribution imprints are compared against ARTEMIS observations.

\section{Computational setup}
We model a one-dimensional slice across the lunar wake as it is convected by the solar wind, so that the simulation's temporal evolution maps to radial distance from the Moon via $r = v_{\mathrm{sw}} t$. This approach follows \citeA{birch2001detailed,birch2001particle}. The simulation domain spans $-L_x/2 \leq x \leq L_x/2$ with $L_x = 60\,d_i$, discretized on $300{,}000$ grid cells, with the interplanetary magnetic field along $+x$. The initial density is $n = n_0$ for $R_l \leq |x| \leq L_x/2$ and $n = 0$ otherwise, where $n_0$ is the solar wind reference density and $R_l = 13.2\,d_i$ is the lunar radius. We use a reduced mass ratio $m_i/m_e = 100$ to keep the computation tractable.

The electron velocity distribution consists of three components (core, halo, and strahl) constructed based on spacecraft observations \cite{vstverak2009radial,maksimovic2000solar}, with densities $0.9n_0$, $0.06n_0$, and $0.04n_0$, respectively. The core is a Maxwellian with thermal velocity $v_{T,c}/c_s = \sqrt{m_i/m_e} = 10$, where $c_s = \sqrt{T_c/m_i}$ is the ion acoustic speed and $T_c = m_e v_{T,c}^2$ is the core temperature. The halo is a Kappa distribution with $v_{T,h} = 2v_{T,c}$ and $\kappa_h = 5$. The strahl is a shifted Kappa distribution with parallel velocity shift $\Delta_s = 2.5\,v_{T,c}$, thermal velocity $v_{T,s} = 2\,v_{T,c}$, and $\kappa_s = 5$. The ion distribution is a Maxwellian with thermal velocity $v_{T,i}/c_s = \sqrt{T_i/T_c} = 0.82$. Functional forms are given in the Supporting Information.

Particles crossing the domain boundaries are removed from the simulation, and new particles are independently injected at both boundaries by sampling from the prescribed solar wind velocity distributions; the injected electron flux differs between the two boundaries due to the unidirectional strahl. To trace particle origins as populations mix within the wake, electrons originating from the left and right sides of the domain are stored in separate arrays throughout the simulation, enabling identification of reflected and penetrating populations in the velocity distribution functions. Radial distance maps to simulation time roughly as $r/R_l = 5.5 \times 10^{-4}\,t\,\omega_{pi}$ \cite{An2025plasma}. Details of the field and particle boundary conditions, damping layers, and time step are given in the Supporting Information.

\section{Results}
\subsection{Two-scale structure}
\begin{figure}[htbp!]
    \centering
    \includegraphics[width=\linewidth]{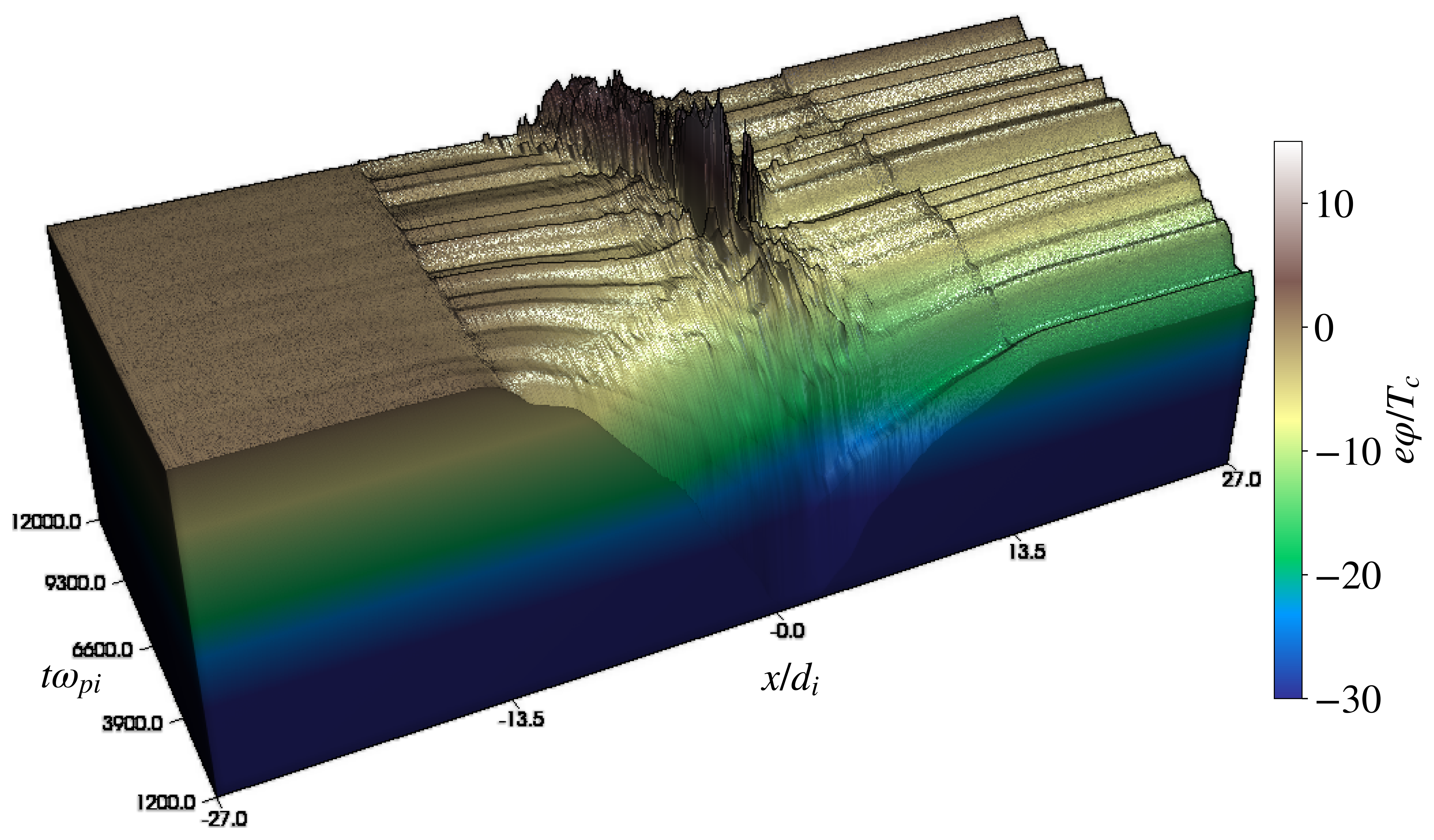}
    \caption{Spatiotemporal evolution of lunar wake potentials. The potential at the left boundary ($x = -27\,d_i$) is set to zero as a reference at all times. The time axis maps to radial distance downstream of the Moon via $r/R_l = 5.5 \times 10^{-4}\,t\,\omega_{pi}$. The volume rendering reveals two distinct spatial scales: the macroscale potential well associated with plasma expansion into the vacuum, and the microscale potential enhancements associated with ion acoustic shocks near the central wake. Small abrupt changes at $x/d_i = \pm 13.2$ mark the lunar surface boundary ($R_l = 13.2\,d_i$).}
    \label{fig:tjoin-pot}
\end{figure}

Figure~\ref{fig:tjoin-pot} shows the spatiotemporal evolution of lunar wake potentials. As electrons, being faster than ions, lead the expansion from the solar wind into the vacuum, an ambipolar electric field develops pointing from the solar wind toward the vacuum. This produces a potential well with its global minimum near the central wake ($x = 0$). Its minimum is shifted toward $x > 0$ due to strahl electrons streaming in the $+x$ direction. The same strahl component also causes the potential on the left side of the wake to exceed that on the right: strahl electrons carry excess flux in $+x$, breaking the symmetry between the two sides. This asymmetry is set by the strahl's field-aligned streaming direction rather than by the IMF polarity, and therefore persists regardless of whether the IMF points toward or away from the Sun. Following the self-similar description of plasma expansion into a vacuum \cite<e.g.,>[]{gurevich1966self,crow1975expansion,denavit1979collisionless}, this potential well has a length scale growing with distance from the Moon as $c_s t$; we refer to this as the \textit{macroscale} potential.

As the two expanding plasma fronts from opposite sides of the wake meet near the central wake, the counter-propagating supersonic ion beams interact strongly, generating ion acoustic shocks that decelerate the beams to subsonic velocities \cite{An2025plasma}. The associated potential enhancements, which increase from upstream to downstream of each shock as plasma density is compressed across the supersonic-to-subsonic transition, span tens of local Debye lengths, much shorter than the macroscale potential well; we refer to these as the \textit{microscale} potentials. By the time the shocks form, the macroscale potential well has grown to a length scale of order $R_l \sim 13\,d_i$, while the microscale potential enhancements span tens of local Debye lengths, $\lambda_{De}$. Their ratio is $R_l/\lambda_{De} = 13\,c/c_s \sim 2.4\times10^4$ in the simulation (see Supporting Information for the derivation), or $\sim 10^5$ using realistic mass ratio and solar wind parameters, underscoring the vast scale separation between the two components of the wake potential.


Because both plasma expansion and shock formation evolve on the ion plasma time scale, much slower than the electron plasma period, electrons are expected to remain in force balance, $-en_eE_x = \partial_x p$, where $p$ is the $(x,x)$ component of the electron pressure tensor. We verify this using the cumulative electric force and pressure difference from the left boundary (see Supporting Information for the integral formulation and full comparison). Force balance holds closely throughout the domain at both spatial scales, including within the shock transition layer, with deviations confined to the abrupt potential changes at the lunar surface boundary ($x = \pm R_l$). This confirms the quasi-static electron response assumed by existing methods that infer wake potentials from electron velocity distributions \cite{halekas2005electrons,xu2019mapping,an2026inferring}, since force balance guarantees that the electron distribution adjusts to the local potential fast enough to be treated as being in equilibrium at each instant.

\subsection{Mediation of energy conversion}\label{sec:energy}
\begin{figure}[htbp!]
    \centering
    \includegraphics[width=\linewidth]{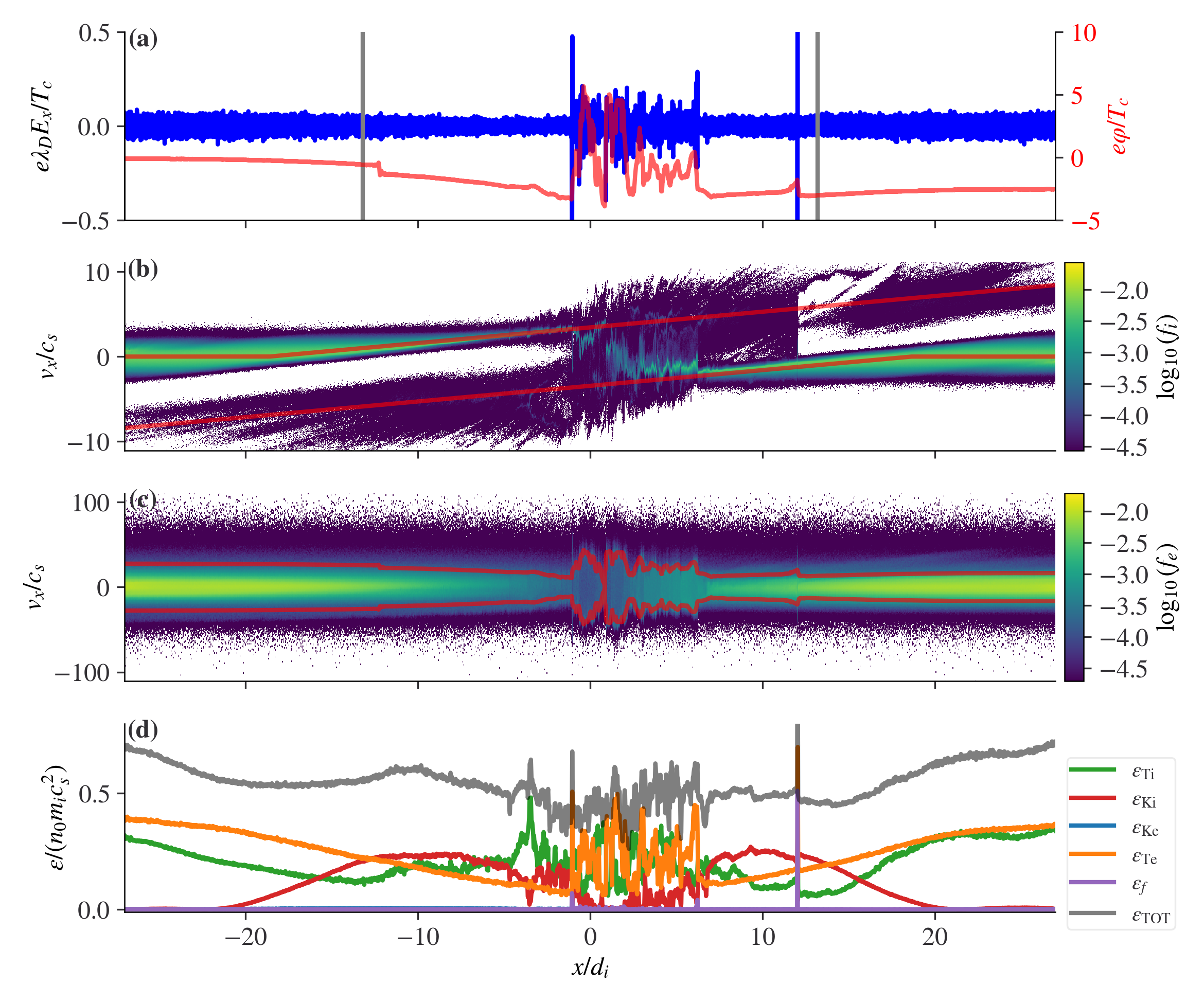}
    \caption{Mediation of energy conversion between ions and electrons by the two-scale electric potential. Simulation data are from $t\omega_{pi} = 10000$. (a) Electric field (blue) and electric potential (red). (b) Ion phase portrait $(x, v_x)$, with color encoding the logarithm of ion phase space density $f_i$. Red lines show the fluid theory prediction $v_x/c_s = 1 + (x \pm R_l)/(c_s t)$. (c) Electron phase portrait $(x, v_x)$, with color encoding the logarithm of electron phase space density $f_e$. Red lines show the separatrix $v_{x,\mathrm{sep}} = \pm\sqrt{(2e/m_e)(\varphi - \varphi_{\min})}$, where $\varphi_{\min}$ is the global minimum potential. (d) Particle and field energy densities. The bulk kinetic and thermal energy densities of species $s$ ($s \in \{e, i\}$) are defined as $\varepsilon_{Ks} = \frac{1}{2}m_s n_s u_s^2$ and $\varepsilon_{Ts} = \frac{1}{2}m_s \int \mathrm{d}v_x\,(v_x - u_s)^2 f_s(x, v_x)$, respectively, where $n_s = \int \mathrm{d}v_x\, f_s(x,v_x)$ and $u_s = \int \mathrm{d}v_x\, v_x f_s(x,v_x)/n_s$. Field energy density is $\varepsilon_f = E_x^2/(8\pi)$. Total energy density (gray) is $\varepsilon_{\mathrm{TOT}} = \sum_s(\varepsilon_{Ks} + \varepsilon_{Ts}) + \varepsilon_f$.}
    \label{fig:energy-conversion}
\end{figure}

Figure~\ref{fig:energy-conversion} shows a snapshot of the electric field, ion and electron phase portraits, and various forms of energy density across the lunar wake, illustrating how the two-scale potential mediates energy transfer between ions and electrons. The electric field itself stores negligible energy, as shown below; its role is not to hold energy but to couple the ion and electron populations, converting energy between bulk kinetic and thermal forms as it passes from one species to the other.

In the expansion region [$-27 < x/d_i < -2$ and $7 < x/d_i < 27$; Figure~\ref{fig:energy-conversion}(a)], the macroscale potential drop, generated by faster electrons leading ahead of ions, simultaneously accelerates ions to supersonic velocities [Figure~\ref{fig:energy-conversion}(b)] and decelerates electrons [Figure~\ref{fig:energy-conversion}(c)]. Mediated by this potential, electron thermal energy is converted into ion bulk kinetic energy [Figure~\ref{fig:energy-conversion}(d)]. The accompanying drop in ion thermal velocity reflects the velocity filter effect: faster ions arrive at the central wake earlier than slower ones, narrowing the local ion velocity distribution.

In the central wake [$-2 < x/d_i < 7$; Figure~\ref{fig:energy-conversion}(a)], ion acoustic shocks generated by two counter-propagating supersonic ion beams simultaneously decelerate ions to subsonic velocities and thermalize them [Figure~\ref{fig:energy-conversion}(b)], while significantly heating electrons [Figure~\ref{fig:energy-conversion}(c)]. Here the microscale potential mediates the reverse conversion, from ion bulk kinetic energy back into ion and electron thermal energies [Figure~\ref{fig:energy-conversion}(d)]. As a result, despite the lower density in the central wake, ion and electron thermal energy densities recover to levels comparable to those in the solar wind, consistent with energy conservation. Throughout this process, the electric field energy remains small compared to particle kinetic and thermal energies, confirming that the field does not itself store the transferred energy: it acts purely as the mediator, enabling energy conversion between bulk kinetic and thermal forms and channeling energy between ions and electrons at both spatial scales.

\subsection{Imprints on particle velocity distributions}
The two-scale lunar wake potential imprints distinct signatures on both ion and electron velocity distributions. On the ion side, the macroscale potential accelerates ions into supersonic beams, while the microscale potential decelerates and thermalizes them --- signatures already discussed in the context of energy conversion (Section~\ref{sec:energy}). On the electron side, the phase space separatrix divides electrons into reflected, penetrating, and nonlinearly trapped populations, encoding information about the underlying electric potentials. These electron signatures are particularly valuable for inferring wake potentials \cite{halekas2005electrons,halekas2014first,xu2019mapping,an2026inferring}, since the potential gradient is typically below the measurement capability of current electric field instruments. Here we characterize the electron velocity distribution imprints in detail using both simulations and ARTEMIS observations.

\begin{figure}[htbp!]
    \centering
    \includegraphics[width=\linewidth]{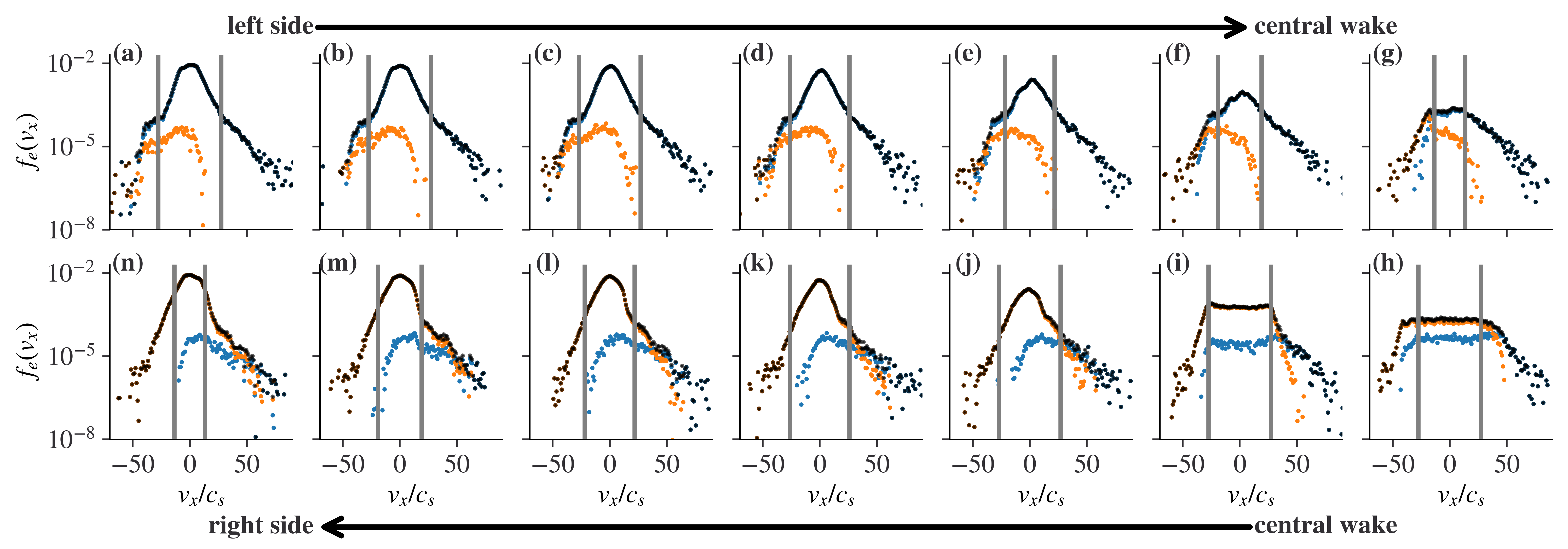}
    \caption{Electron parallel velocity distributions at selected locations across the wake at $t\omega_{pi} = 10000$. Blue and orange show electrons originating from the left and right sides of the wake, respectively; black shows the total phase space density. Vertical gray lines mark the separatrix velocities $v_{x,\mathrm{sep}} = \pm\sqrt{(2e/m_e)(\varphi - \varphi_{\min})}$. Panels (a)--(n) correspond to locations $x/d_i = -26, -22, -18, \ldots, -2, 2, \ldots, 18, 22, 26$, ordered from left to right as indicated by the arrows.}
    \label{fig:fvpara}
\end{figure}

Figure~\ref{fig:fvpara} shows $14$ electron parallel velocity distributions spanning from the left side of the wake through the central wake to the right side, at the same time instant as Figure~\ref{fig:energy-conversion}. Electrons originating from the two sides of the wake are tracked separately in the simulation, enabling identification of reflected and penetrating populations. Strahl electrons, centered at $v_x = 20\,c_s$ with thermal velocity $25\,c_s$ and $\kappa = 5$, penetrate in significant numbers from the left to the right side, forming the high-energy tail ($v_x/c_s \gtrsim 40$) observed on the right side.

In the central wake, microscale potential enhancements heat the electron velocity distributions; these are further flattened through nonlinear Landau resonance, forming flat-top distributions. A beam feature near the negative separatrix velocity originates in the central wake and propagates toward the left side. We attribute this beam to reflection of strahl electrons near the positive separatrix velocity, which carries comparable phase space density; the contribution of right-side penetrating electrons to this beam is secondary. This interpretation suggests an alternative explanation to \citeA{halekas2014first}, who attribute the beam to penetrating electrons traveling opposite to the strahl. We note that in a symmetric electron velocity distribution, such a beam would not arise, as neither reflected nor penetrating electrons would have enhanced phase space density near the separatrix in one direction relative to the other.

\subsection{ARTEMIS observations}
We substantiate the simulation results with ARTEMIS observations. The ARTEMIS mission consists of two spacecraft (P1 and P2) in equatorial lunar orbits \cite{angelopoulos2014artemis}, regularly crossing the lunar wake. This two-spacecraft configuration is well suited to studying wake refilling: one spacecraft traverses the wake while the other monitors the ambient solar wind, providing simultaneous upstream context. We analyze plasma data from the electrostatic analyzer at $\sim 4$\,s resolution \cite{mcfadden2008themis-a4f} and magnetic field data from the fluxgate magnetometer at $\sim 0.0625$\,s resolution \cite{auster2008themis-f2e}.

\begin{figure}[htbp!]
    \centering
    \includegraphics[width=\linewidth]{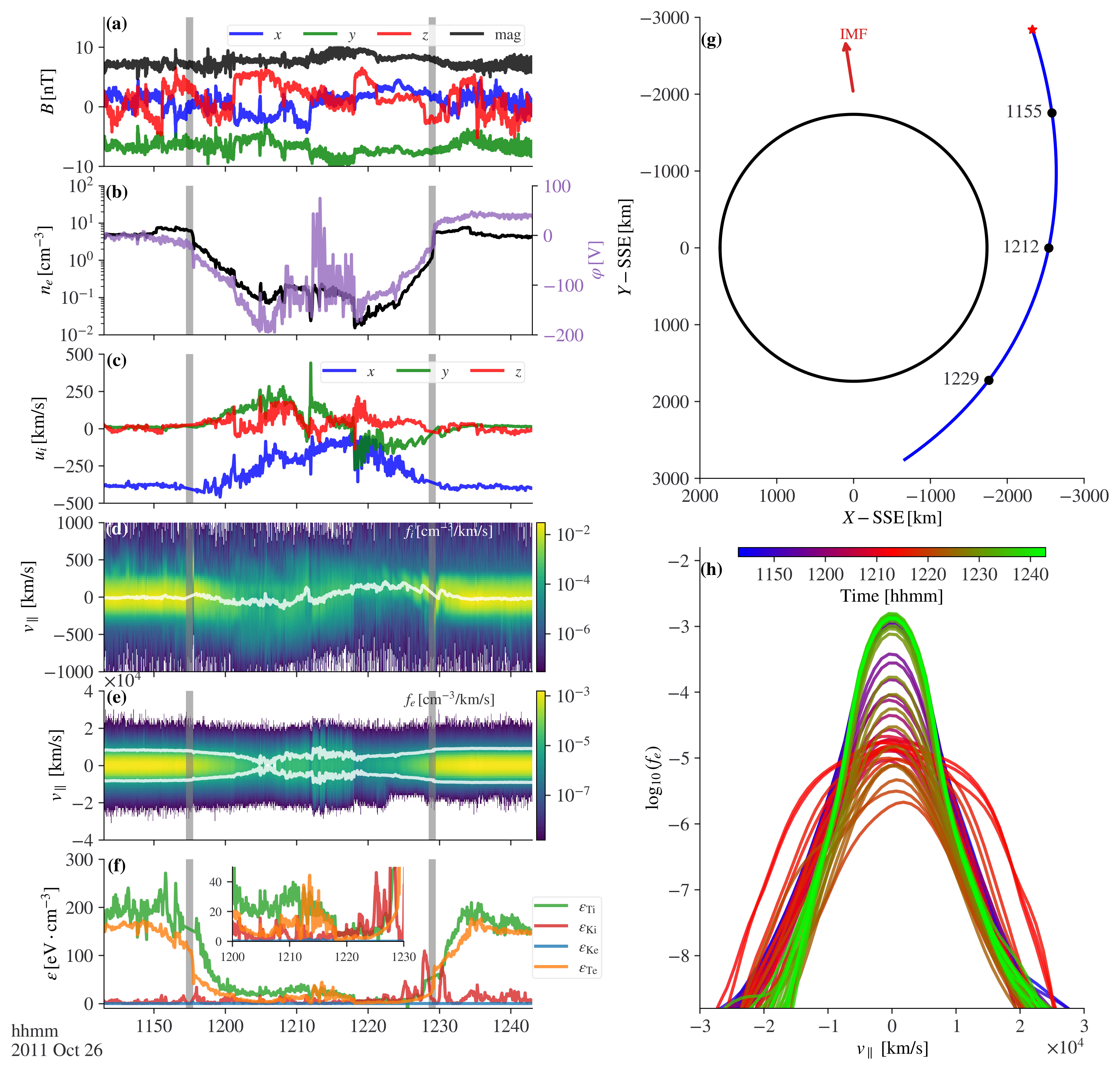}
    \caption{ARTEMIS P1 observations of the lunar wake crossing on October~26, 2011. The two vertical gray bars in each panel indicate the entry and exit points of the wake. (a) Magnetic field components and magnitude. (b) Electron density (black) and inferred electric potential (purple, right axis). (c) Ion bulk flow velocity components. (d) Ion phase space density as a function of parallel velocity; the white line marks the parallel ion bulk flow velocity. The anomalies in the ion velocity distributions at the wake entry/exit points may be unphysical. (e) Electron phase space density as a function of parallel velocity; the white curve marks the separatrix, defined as the locus of points with $v_\parallel = 0$ at the location of the global potential minimum. (f) Thermal and bulk kinetic energy densities for ions and electrons. (g) Spacecraft trajectory in the $X$-$Y$ plane in SSE coordinates, showing the lunar surface (black circle), the trajectory within the analyzed interval (blue), the mean IMF direction (red arrow), and the times of wake boundary crossings and central wake passage (black dots). (h) Electron parallel velocity distributions at selected times during the wake crossing, colored by time.}
    \label{fig:artemis}
\end{figure}

As a case study, we examine a single ARTEMIS wake crossing on October~26, 2011, between $\sim 11{:}50$ and $\sim 12{:}40$~UT, during which P1 traversed the lunar wake while P2 remained in the solar wind. Both probes recorded the IMF predominantly in the $-y$ direction in Selenocentric Solar Ecliptic (SSE) coordinates [Figure~\ref{fig:artemis}(a); P2 not shown]. Despite significant magnetic discontinuities in the $x$ and $z$ components, this predominantly $-y$ orientation provides favorable conditions for observing parallel refilling of the lunar wake in the equatorial plane. A density void was present in the central wake, with density dropping by two orders of magnitude relative to the ambient solar wind (from $\sim 10\,\mathrm{cm}^{-3}$ to $\sim 0.1\,\mathrm{cm}^{-3}$) [Figures~\ref{fig:artemis}(b) and \ref{fig:artemis}(g)]. Density compressions were observed near $12{:}07$ and $12{:}18$~UT, likely associated with counter-streaming ion flows transitioning from supersonic ($\sim \pm 200$\,km/s in $y$) to subsonic velocities across ion acoustic shocks [Figure~\ref{fig:artemis}(c)]; the local upstream ion acoustic speed is $98$\,km/s, compared to $59$\,km/s in the solar wind. This single event serves as a proof of concept; a statistical survey across many crossings and solar wind conditions is left for future work.

The presence of ion acoustic shocks and the associated energy conversion are further supported by particle velocity distribution measurements. Figures~\ref{fig:artemis}(d) and \ref{fig:artemis}(e) show ion and electron phase space densities, respectively, as a function of parallel velocity at each time instant during the wake crossing. The acceleration of ions into counter-propagating supersonic beams toward the wake, followed by their deceleration and thermalization in the central wake, is evident in Figure~\ref{fig:artemis}(d). Part of the ion heating is attributable to magnetic discontinuities, particularly prior to the interaction of the two ion beams (e.g., between $11{:}55$ and $12{:}10$~UT). In the region where the ion beams decelerate, electrons are significantly heated and form flat-top distributions [Figures~\ref{fig:artemis}(e) and \ref{fig:artemis}(h)], consistent with the simulated ion and electron phase space densities [Figures~\ref{fig:energy-conversion}(b), \ref{fig:energy-conversion}(c), and \ref{fig:fvpara}(g-i)].

To understand the energy conversion between ions and electrons, we calculate the thermal ($\varepsilon_{Ts}$) and bulk kinetic ($\varepsilon_{Ks}$) energy densities for the two species ($s \in \{i, e\}$) in Figure \ref{fig:artemis}(f). Because of additional ion heating by magnetic discontinuities on the entry side of the wake [before $\sim 12{:}10$~UT; Figures~\ref{fig:artemis}(a) and \ref{fig:artemis}(d)], we focus on the exit side after $\sim 12{:}10$~UT, where the IMF is relatively steady. Toward the wake, electron and ion thermal energy densities decrease exponentially while the ion bulk kinetic energy density increases. In the central wake, around $12{:}12$~UT, both electron and ion thermal energy densities are enhanced.

The parallel ion bulk flow velocity decreases across the ion acoustic shocks by $\Delta u_{i,\parallel} \approx 150$~km/s [Figure~\ref{fig:artemis}(d)], sufficient to account for the increase in electron thermal velocity, $\Delta v_{Te,\parallel} \approx 3000\,\mathrm{km/s} \lesssim \sqrt{m_i/m_e}\,\Delta u_{i,\parallel} \approx 6000$~km/s [Figures~\ref{fig:artemis}(e) and \ref{fig:artemis}(h)].

To quantify this energy conversion, we infer electric potentials from electron phase space densities following the Hamiltonian inversion method of \citeA{an2026inferring}, shown in Figure~\ref{fig:artemis}(b). This method exploits the conservation of the Hamiltonian along individual electron trajectories to invert the observed phase space density for the underlying electric potential. The maximum macroscale potential drop from the ambient solar wind to the wake is $\Delta\varphi_{\max} = 100$~V, corresponding to an ion bulk flow velocity $\Delta u_{\max} = \sqrt{2e\Delta\varphi_{\max}/m_i} = 140$~km/s. This closely matches the independently measured parallel ion bulk velocity of $150$~km/s [Figure~\ref{fig:artemis}(d)], a direct observational confirmation that the electron-inferred potential correctly predicts the ion acceleration it is understood to drive.

The microscale potential enhancement across the shock, $\Delta\varphi_{\mathrm{shock}} \sim 100$~V, decelerates the supersonic ion flow while simultaneously heating electrons: $\Delta v_{Te,\parallel} = \sqrt{2e\Delta\varphi_{\mathrm{shock}}/m_e} = \sqrt{m_i/m_e}\,\Delta u_{\max} \approx 6000$~km/s, consistent with the measured increase in electron thermal velocity [Figures~\ref{fig:artemis}(e) and \ref{fig:artemis}(h)]. Together, these results substantiate the simulated picture: the macroscale potential accelerates counter-propagating supersonic ion beams, while the microscale potential mediates the conversion of ion bulk kinetic energy back into electron and ion thermal energy.

\section{Conclusion}
Using PIC simulations substantiated by ARTEMIS observations, we have characterized the critical role of electric potentials in the lunar wake. The wake potential exhibits two distinct spatial scales: a macroscale component associated with self-similar plasma expansion into the vacuum, and a microscale component associated with ion acoustic shocks generated by counter-propagating supersonic ion beams in the central wake. We verify that electrons remain in force balance throughout the wake, with the electric force balanced by the electron pressure gradient at both spatial scales. The two-scale potential mediates a sequential energy conversion: the macroscale potential converts electron thermal energy into ion bulk kinetic energy, while the microscale potential converts it back into electron and ion thermal energies. Both scales imprint distinct signatures on ion and electron velocity distributions (e.g., ion beams and electron flat-top distributions), reproduced in simulations and confirmed in ARTEMIS observations.

This study provides a physical foundation for inferring lunar wake electric potentials from electron phase space density measurements \cite{an2026inferring}. The verification of electron force balance justifies the Vlasov equilibrium assumption underlying existing inference methods, while the characterization of velocity distribution imprints identifies the physical origin of the electron signatures these methods exploit. The electric field plays a central role in this energy transfer despite its energy density being small compared to particle kinetic and thermal energies --- a result that may extend to other airless body wakes where parallel refilling dominates.

Future work could extend the statistical maps of lunar wake electric potentials \cite{halekas2005electrons,xu2019mapping} to include the microscale shock enhancements identified here, accounting for the left-right potential asymmetry and mixed electron populations near the shock. The present findings are expected to hold locally in higher dimensions, since the parallel force balance and energy conversion mechanisms are set by field-aligned dynamics; however, extending the framework beyond one dimension would reveal how the two-scale structure couples to cross-field plasma expansion, governing the full three-dimensional structure of the lunar wake.

\appendix
\section{Electron velocity distributions in the solar wind}
The electron velocity distribution function (VDF) consists of three components:
\begin{linenomath*}
    \begin{align}
        f = f_c + f_h + f_s,
    \end{align}
\end{linenomath*}
where $f_c$, $f_h$, and $f_s$ are the VDFs of the core, halo, and strahl, respectively. We adopt the functional forms and parameters of the three components from \citeA{vstverak2009radial} and \citeA{maksimovic2000solar}, constructed from statistical spacecraft observations. The analytical expressions for $f_c$, $f_h$, and $f_s$ are included below for completeness.

The core electron VDF is
\begin{linenomath*}
    \begin{align}
        f_c = A_c \exp\left[-\frac{v_\perp^2 + (v_\parallel - \Delta_c)^2}{2 v_{\mathrm{Tc}}^2}\right] ,
    \end{align}
\end{linenomath*}
where $v_\perp$ and $v_\parallel$ are the perpendicular and parallel velocities with respect to the background magnetic field, respectively, $v_{\mathrm{Tc}}$ is the core thermal velocity, and $\Delta_c$ is the drift velocity in the plasma frame. The normalization factor $A_c$ is
\begin{linenomath*}
    \begin{align}
        A_c = n_c \left(\frac{1}{2 \pi v_{\mathrm{Tc}}^2}\right)^{3/2} ,
    \end{align}
\end{linenomath*}
where $n_c$ is the core number density.

The halo electron VDF is
\begin{linenomath*}
    \begin{align}
        f_h = (1 - f_{h,ft}) f_{h,\kappa} ,
    \end{align}
\end{linenomath*}
where $f_{h,\kappa}$ is the Kappa function and $f_{h,ft}$ is a flat-top function that truncates the inner part of the Kappa function overlapping with the thermal core. The Kappa function is given by
\begin{linenomath*}
    \begin{align}\label{eq:fhkappa}
        f_{h,\kappa} = A_h \left(1 + \frac{v_\perp^2 + v_\parallel^2}{(2 \kappa_h - 3) v_{\mathrm{Th}}^2}\right)^{-\kappa_h - 1} ,
    \end{align}
\end{linenomath*}
where $\kappa_h$ describes the ``hardness'' of the halo and $v_{\mathrm{Th}}$ describes its velocity ``spread.'' The normalization factor $A_h$ is
\begin{linenomath*}
    \begin{align}
        A_h = n_{h\kappa} \left(\frac{1}{\pi(2\kappa_h-3)v_{\mathrm{Th}}^2}\right)^{3/2} \frac{\Gamma(\kappa_h + 1)}{\Gamma(\kappa_h - 0.5)} ,
    \end{align}
\end{linenomath*}
where $n_{h\kappa}$ is the zeroth-order moment of $f_h$ and $\Gamma(\cdot)$ is the Gamma function. Because of the truncation, $n_{h\kappa}$ does not equal the exact halo number density. The flat-top function is
\begin{linenomath*}
    \begin{align}
        f_{h,ft} = \left[1 + \left(\frac{v_\perp^2 + (v_\parallel - \Delta_c)^2}{2 \delta v_{\mathrm{Tc}}^2}\right)^p\right]^{-q} ,
    \end{align}
\end{linenomath*}
where $\delta$ controls the width of the flat top, and the fixed parameters $p=10$ and $q=1$ make the edge of the flat top fall rapidly to zero.

The strahl electron VDF is
\begin{linenomath*}
    \begin{align}
        f_s = A_s \left(1 + \frac{v_\perp^2 + D (v_\parallel-\Delta_s)^2}{(2 \kappa_h - 3) v_{\mathrm{Ts}}^2}\right)^{-\kappa_s - 1} ,
    \end{align}
\end{linenomath*}
with
\begin{linenomath*}
    \begin{align}
        \begin{cases}
            D=1, \hspace{10pt} \mathrm{for}\, v_\parallel \geq \Delta_s , \\
            D=\Theta, \hspace{10pt} \mathrm{for}\, v_\parallel < \Delta_s ,
        \end{cases}
    \end{align}
\end{linenomath*}
where $\kappa_s$ and $v_{\mathrm{Ts}}$ have meanings analogous to those in Equation \eqref{eq:fhkappa}, $\Delta_s$ is the strahl drift in the plasma frame, and $\Theta$ is fixed at $10$ to truncate the VDF sunward of $v_\parallel < \Delta_s$. The normalization factor is
\begin{linenomath*}
    \begin{align}
        A_s = n_{s} \frac{2 \sqrt{\Theta}}{\sqrt{\Theta+1}} \left(\frac{1}{\pi(2\kappa_h-3)v_{\mathrm{Ts}}^2}\right)^{3/2} \frac{\Gamma(\kappa_s + 1)}{\Gamma(\kappa_s - 0.5)} ,
    \end{align}
\end{linenomath*}
where $n_s$ is the strahl number density.

\begin{figure}[htpb]
    \centering
    \includegraphics[width=\linewidth]{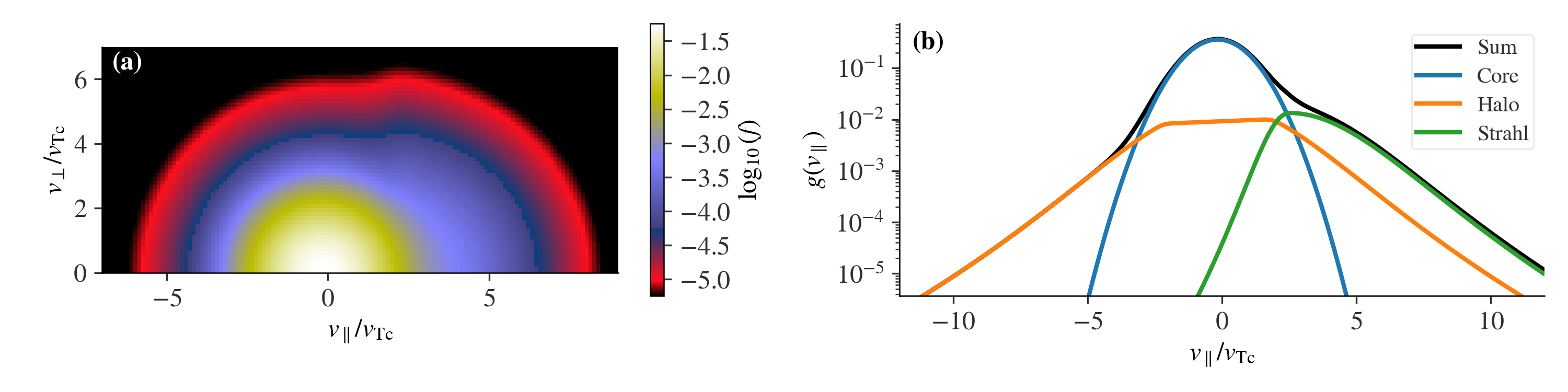}
    \caption{Electron velocity distribution function. (a) Electron VDF in $(v_\parallel, v_\perp)$ space. (b) Reduced electron VDF $g(v_\parallel) = \int 2\pi v_\perp\, f(v_\perp, v_\parallel)\, \mathrm{d}v_\perp$ for the core (blue), halo (orange), and strahl (green) components, and their sum (black).}
    \label{fig:eVDF}
\end{figure}

Table~\ref{table:eVDF-params} lists the parameters of the three-component electron velocity distribution function used in the simulation. Figure~\ref{fig:eVDF} shows the electron VDF in $(v_\parallel, v_\perp)$ space and the reduced VDF, $g(v_\parallel) = \int 2\pi v_\perp\, f(v_\perp, v_\parallel)\, \mathrm{d}v_\perp$, which clearly reveal the non-Maxwellian suprathermal tail and the $v_\parallel$ asymmetry caused by the strahl.

\begin{table}[htpb]
\centering
\caption{Parameters of the three-component electron velocity distribution function used in the simulation. Velocities are normalized to the core thermal velocity $v_{\mathrm{Tc}}$, densities to the solar wind reference density $n_0$. The core thermal velocity is $v_{\mathrm{Tc}} = 10\,c_s$, where $c_s$ is the ion acoustic speed; the core drift $\Delta_c$ is set to balance the net electron flux carried by the strahl, ensuring zero net current.\label{table:eVDF-params}}
\begin{tabular}{llc}
\hline
Component & Parameter & Value \\
\hline
Core   & $v_{\mathrm{Tc}}$   & $1.0\,v_{\mathrm{Tc}}$ \\
       & $n_c$               & $0.9\,n_0$ \\
       & $\Delta_c$          & $-0.160\,v_{\mathrm{Tc}}$ \\
\hline
Halo   & $v_{\mathrm{Th}}$   & $2.0\,v_{\mathrm{Tc}}$ \\
       & $\kappa_h$          & $5.0$ \\
       & $n_{h\kappa}$       & $0.06\,n_0$ \\
       & $\delta$            & $2.0$ \\
\hline
Strahl & $v_{\mathrm{Ts}}$   & $2.0\,v_{\mathrm{Tc}}$ \\
       & $\Delta_s$          & $2.5\,v_{\mathrm{Tc}}$ \\
       & $\kappa_s$          & $5.0$ \\
       & $n_s$               & $0.04\,n_0$ \\
       & $\Theta$            & $10.0$ \\
\hline
\end{tabular}
\end{table}

\section{Particle injection boundary condition}
At each of the two domain boundaries ($x = x_0$ and $x = x_1$), particles are injected to represent the unperturbed solar wind flowing into the simulation domain. Two aspects of the algorithm are described below: (1) computation of the injection flux for each species, and (2) sampling of injected particle velocities from the prescribed boundary velocity distributions.

\noindent\textit{Injection flux}---The injection flux determines the mean number of particles injected per boundary cell per timestep. For ions, which follow a drifting Maxwellian, the flux across the boundary has the closed-form expression
\begin{linenomath*}
    \begin{align}
        \Phi_i = \frac{n_i v_{\mathrm{Ti}}}{\sqrt{2}}\left[\frac{\exp(-\zeta^2)}{\sqrt{\pi}} + \zeta\left(1 + \mathrm{erf}(\zeta)\right)\right] ,
    \end{align}
\end{linenomath*}
where $\zeta \equiv u_i/(\sqrt{2}\,v_{\mathrm{Ti}})$ is the drift velocity normalized to the ion thermal velocity, and the sign of $\zeta$ is set according to the boundary (left or right) so that only the inward-directed flux is injected.

For electrons, whose three-component (core, halo, strahl) velocity distribution has no simple closed form for the flux, the flux is instead computed by direct numerical integration of the distribution function over the half of velocity space directed into the domain:
\begin{linenomath*}
    \begin{align}
        \Phi_e = 2\pi \int_0^\infty \mathrm{d}v_\perp\, v_\perp \int_{0}^{\infty} \mathrm{d}v_x\, v_x\, f_e(v_x, v_\perp) ,
    \end{align}
\end{linenomath*}
with $v_x$ measured into the domain at each boundary. This integral is evaluated once, at the start of the simulation, using a discretized trapezoidal sum over a fine two-dimensional velocity grid.

\noindent\textit{Injection algorithm}---At each timestep, an injection accumulator $b$ is incremented at every boundary cell by the flux times the timestep, $b \mathrel{+}= \Phi\,\Delta t$. The integer part of $b$ gives the number of particles to inject in that cell during that step, and the accumulator is reduced by this integer, retaining the fractional remainder for the next step. This scheme ensures that, on average, the injection rate matches $\Phi$ even though $\Phi\,\Delta t$ is typically much less than one particle per cell per step.

For each particle to be injected, the velocity is sampled by rejection sampling against the appropriate boundary distribution function. The overall procedure is summarized in the following pseudocode:

\begin{verbatim}
for each boundary (left, right):
    for each species s:
        accumulate injection count: b[s] += Phi[s] * dt
        n_inject = floor(b[s])
        b[s] -= n_inject

        repeat n_inject times:
            if s is electron:
                # rejection sampling from 3-component (core+halo+strahl) VDF
                repeat:
                    draw candidate (v_par, v_perp) uniformly
                        over [0, 12 vTc] x [0, 12 vTc]  (inward half-space)
                    draw f_rand uniformly over [0, f_max]
                    evaluate f_electron(v_par, v_perp)
                until f_rand <= f_electron(v_par, v_perp)
                draw gyrophase angle theta uniformly over [0, 2*pi]
                (v_y, v_z) = (v_perp*cos(theta), v_perp*sin(theta))

            if s is ion:
                draw (v_y, v_z) from Maxwellian with thermal velocity vTi
                repeat:
                    draw candidate v_par uniformly over [0, 5 vTi]
                        (inward half-space)
                    draw f_rand uniformly over [0, 1]
                    evaluate f_ion(v_par) = exp(-0.5 * (v_par/vTi)^2)
                until f_rand <= f_ion(v_par)

            place particle at boundary with random position
                within the transverse cell area, age = 0
            inject_particle(x_boundary, y, z, v_par, v_y, v_z)
\end{verbatim}

For electrons, the inward-directed parallel velocity $v_x$ and the perpendicular speed $v_\perp$ are sampled jointly by rejection against the full three-component distribution $f_e = f_c + f_h + f_s$ (main text, Equations~2--7), so that the injected population directly reflects the core, halo, and strahl components together, including the strahl-driven asymmetry between the two boundaries. The gyrophase angle is then drawn uniformly to distribute the perpendicular velocity between the $v_y$ and $v_z$ components. For ions, the two transverse velocity components are drawn directly from a Maxwellian, while the inward parallel component is drawn by rejection sampling against a half-Maxwellian to enforce inward-only flow.

This procedure guarantees that particles injected at each boundary reproduce, in a statistical sense, the prescribed solar wind velocity distributions given in the main text, while enforcing that only inward-propagating particles are added to the simulation.

\section{Field boundary conditions and numerical parameters}
For electromagnetic fields, we use absorbing boundary conditions with charge and current densities accumulated over partial boundary voxels. Two damping layers of thickness $3\,d_i$ at each boundary suppress electric field fluctuations by a factor of $100$, effectively preventing field reflections from the boundaries. The time step $\Delta t = 0.00198\,\omega_{pe}^{-1}$ satisfies the Courant condition.

\section{Macroscale-to-microscale length-scale ratio}
The macroscale potential well has a length scale of order the lunar radius, $R_l \sim 13\,d_i$, where $d_i = c/\omega_{pi}$ is the ion inertial length. The microscale potential enhancements at the ion acoustic shocks span tens of local Debye lengths, $\lambda_{De}$. Using $v_{T,c}/c_s = \sqrt{m_i/m_e}$, the local Debye length satisfies $\lambda_{De} = v_{T,c}/\omega_{pe} = c_s/\omega_{pi}$, so that
\begin{linenomath*}
    \begin{align}
        \frac{R_l}{\lambda_{De}} = \frac{13\,(c/\omega_{pi})}{c_s/\omega_{pi}} = 13\,\frac{c}{c_s} = 13\sqrt{\frac{m_i}{m_e}}\sqrt{\frac{m_e c^2}{T_c}} \approx 2.4\times10^4 ,
    \end{align}
\end{linenomath*}
using the reduced mass ratio $m_i/m_e = 100$ and the simulation's normalized speed of light. For a realistic mass ratio ($m_i/m_e \approx 1836$) and typical solar wind electron temperature, the same estimate gives $R_l/\lambda_{De} \sim 10^5$.

\section{Electron force balance verification}

Both plasma expansion into the vacuum and ion acoustic shock formation evolve on the ion plasma time scale, much slower than the electron plasma period on which electrons respond to electric fields. Electrons are therefore expected to maintain force balance on the slow time scale. We test this by evaluating
\begin{linenomath*}
    \begin{align}
        -e n_e E_x - \partial_x p = 0,
    \end{align}
\end{linenomath*}
where $e$ is the elementary charge, $n_e$ is the electron density, and $p$ is the $(x,x)$ component of the electron pressure tensor. To avoid noise amplification from the spatial derivative $\partial_x p$, we work with the integral form:
\begin{linenomath*}
    \begin{align}
        p(x) - p(x_L) = -\int_{x_L}^x \mathrm{d}x'\, e n_e E_x,
    \end{align}
\end{linenomath*}
defining $F_p(x) \equiv p(x) - p(x_L)$ as the cumulative pressure difference from the left boundary and $F_E(x) \equiv -\int_{x_L}^x \mathrm{d}x'\, e n_e E_x$ as the cumulative electric force. Force balance is satisfied locally wherever $F_E - F_p$ is spatially constant, since $\partial_x(F_E - F_p) = 0$ recovers the differential form.

\begin{figure}[htpb]
    \centering
    \includegraphics[width=\linewidth]{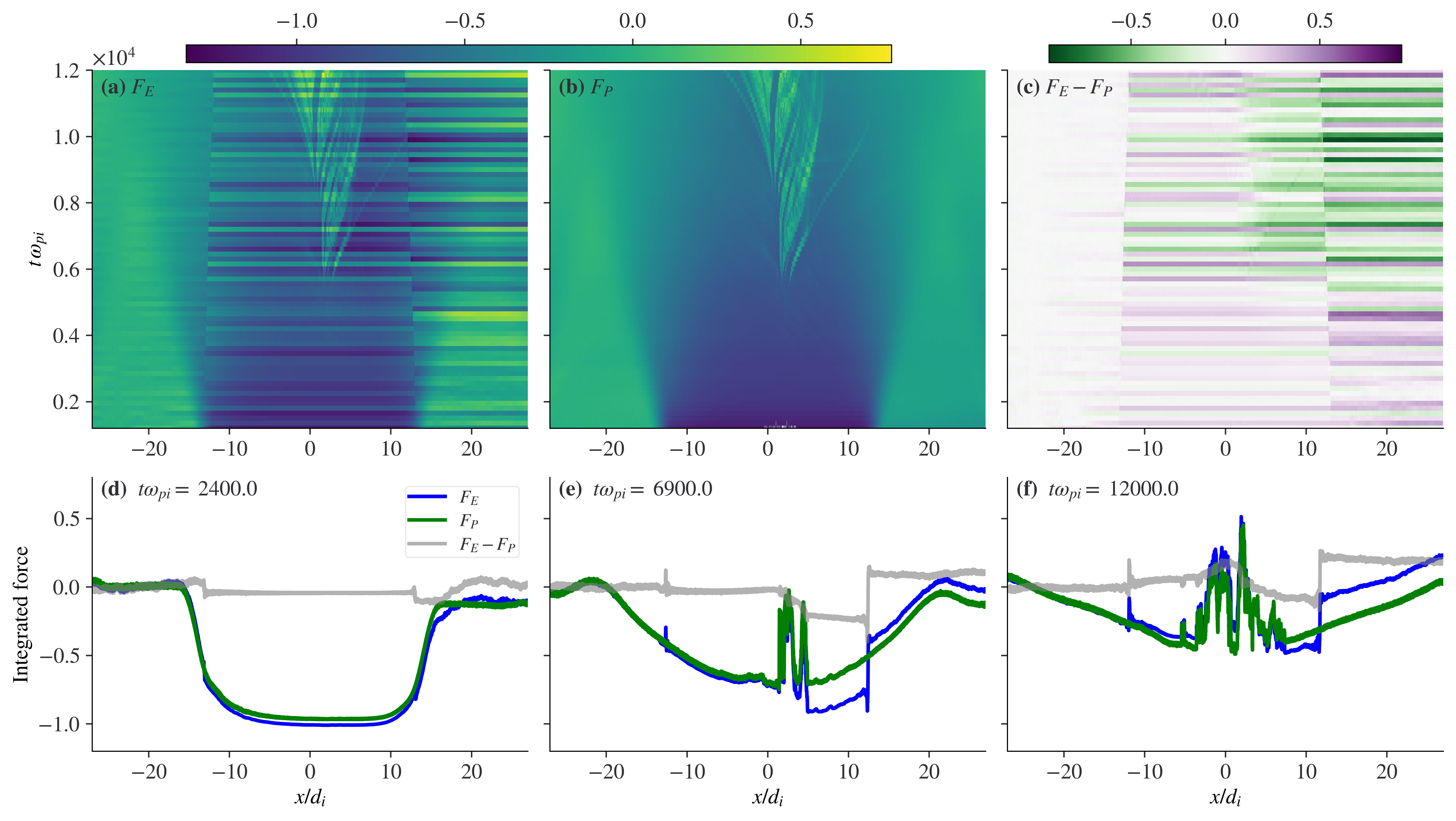}
    \caption{Comparison between the cumulative electric force $F_E$ and the cumulative pressure force $F_p$. (a)--(c) Spatiotemporal evolution of $F_E$, $F_p$, and $F_E - F_p$, respectively. (d)--(f) Spatial profiles of $F_E$ (blue), $F_p$ (green), and $F_E - F_p$ (gray) at $t\omega_{pi} = 2400$, $6900$, and $12000$, respectively. Both $F_E$ and $F_p$ have dimensions of energy density and are normalized to $n_0 T_c$.}
    \label{fig:balance}
\end{figure}

Figure~\ref{fig:balance} shows $F_E$, $F_p$, and their difference. The spatiotemporal evolution of $F_p$ closely resembles that of $\varphi$ (main text Figure~1 and Figure~\ref{fig:balance}(b)): both exhibit the same two-scale structure, with a macroscale component associated with plasma expansion and a microscale component associated with ion acoustic shocks. $F_E$ matches $F_p$ across most of the domain, with deviations confined to sites of abrupt potential variation at $x = \pm R_l = \pm 13.2\,d_i$ (Figures~\ref{fig:balance}(a) and \ref{fig:balance}(c)). Notably, force balance remains valid within the shock transition layer itself, not only in the smoother regions upstream and downstream, indicating that electrons adjust quasi-statically to the electric potential even across the sharp density and field gradients at the shock front.

Line plots at three representative times confirm this: during the pure expansion phase before shock formation ($t\omega_{pi} = 2400$, Figure~\ref{fig:balance}(d)), at the onset of shock formation ($t\omega_{pi} = 6900$, Figure~\ref{fig:balance}(e)), and at the end of the simulation ($t\omega_{pi} = 12000$, Figure~\ref{fig:balance}(f)), $F_E$ and $F_p$ agree closely throughout the domain. Electron force balance $-en_eE_x = \partial_x p$ therefore holds at both spatial scales, with the electric field amplitude differing substantially between the macroscale and microscale components.

%



%
%

\section*{Open Research Section}
The data product and associated Jupyter notebooks used in the analysis are available via \citeA{an2026lunarwakedata} [\url{https://tinyurl.com/wake-pot-struct}].

\acknowledgments
This work was supported by NASA contract NAS5-02099 and NASA grant NO.~80NSSC22K1634. Shaosui Xu gratefully acknowledges support from NASA's Lunar Data Analysis Program (LDAP), grant No.~80NSSC25K7047. The work by Ferdinand Plaschke was financially supported by the German Center for Aviation and Space (DLR) under contract 50 OC 2201. Data access and processing was done using SPEDAS V6.1 \cite{angelopoulos2019space-8b9}. We would like to acknowledge high-performance computing support from Derecho (\url{https://doi.org/10.5065/qx9a-pg09}) provided by NCAR's Computational and Information Systems Laboratory, sponsored by the National Science Foundation \cite{derecho}.

%
%

\bibliography{ref_lunar,references}

%
%
%
%
%

\end{document}